%% file: paper.tex
\documentclass{article}
\usepackage{spconf,amsmath,graphicx, amssymb, cite}

\title{A General Framework for Learning Procedural \\ Audio Models of Environmental Sounds}
\name{
Danzel Serrano and Mark Cartwright
}
\address{
New Jersey Institute of Technology, Newark, NJ, USA \\
\begin{small}
\texttt{ds867@njit.edu}
\end{small}
}
\begin{document}
\maketitle
\begin{abstract}

This paper introduces the \textbf{Pro}cedural (audio) \textbf{V}ariational auto\textbf{E}ncoder (\textit{ProVE}) framework as a general approach to learning Procedural Audio \textbf{PA} models of environmental sounds with an improvement to the realism of the synthesis while maintaining provision of control over the generated sound through adjustable parameters. The framework comprises two stages: \textit{(i) Audio Class Representation}, in which a latent representation space is defined by training an audio autoencoder, and \textit{(ii) Control Mapping}, in which a joint function of static/temporal control variables derived from the audio and a random sample of uniform noise is learned to replace the audio encoder. We demonstrate the use of \textbf{ProVE} through the example of footstep sound effects on various surfaces. Our results show that \textbf{ProVE} models outperform both classical \textbf{PA} models and an adversarial-based approach in terms of sound fidelity, as measured by Fréchet Audio Distance (FAD), Maximum Mean Discrepancy (MMD), and subjective evaluations, making them feasible tools for sound design workflows.
\end{abstract}


\section{Introduction}
\label{secintro}

The job of the sound designer is to record, synthesize, and process audio stimuli for immersive mediums such as film, video games, and virtual environments in such a way that complements visual or other stimuli. The current paradigm typically involves scanning through extensive audio databases for a set of recorded sounds of a sound class $C$ and preprocessing them to fit a project's needs \cite{liljedahl2010methods}. This process is time-consuming and may result in a lack of realism if suitable sounds cannot be found or processed. Storage limitations may also be a problem, arising from the need for a large number of distinct recordings for a highly variable sound class, especially in interactive media such as video games and virtual environments where sound events are influenced by the user's context.

In mitigating the aforementioned problems, Procedural audio (\textbf{PA}) is a different approach to traditional sound design where a sound class is modeled as a mathematical process \cite{farnell2007introduction}. This results in models which reduce storage limitations, have greater control and expressiveness, and have the capacity to create unique auditory experiences. \textbf{PA} models differ from physical modeling synthesis \cite{smith021223physical}, which render sounds through simulations in computational physics. These physics-based models can produce realistic and dynamic sounds, but are computationally expensive and require a significant amount of domain knowledge to develop. On the other hand, \textbf{PA} models are simpler and more computationally efficient, using algorithms to generate sound based on static and temporal control variables, usually accompanied by random noise to span variations in synthesis. Despite its advantage in computational efficiency, current classical \textbf{PA} models still synthesize sounds of lower quality compared to using real samples or physical modeling synthesis --- a primary reason why they are not yet in standard use in sound design \cite{farnell2007introduction, b_ottcher2013current}. 

The state-of-the-art for enhancing sound synthesis quality involves data-driven neural audio synthesis, the subset of deep learning techniques for generative audio. Neural audio synthesis has been extensively researched for speech synthesis \cite{wang2017tacotron, shen2017natural, ren2020fastspeech} and music generation \cite{mittal2021symbolic, dhariwal2020jukebox}, but its application to specific environmental sounds or Foley synthesis for immersive media has only recently been investigated \cite{choi2022proposal2}. Furthermore, the topic of controllability in these models is well established in speech \cite{chen2022controlvc} and music \cite{mittal2021symbolic, dhariwal2020jukebox, wu2021midi} domains, but not yet for environmental sounds. Recently a paper used generative adversarial approaches \cite{kong2020hifi, donahue2019adversarial} to model the environmental sound class of footsteps on various surface materials \cite{comunita2021neural}. Although the model results in a sample space of high-fidelity footstep sounds, its controllability is limited to the surface material selection and its expressivity doesn't account for temporal context, such as locomotion speed or the extent of force from the foot to the ground.

\begin{figure}[htbp]
\begin{minipage}[b]{1.0\linewidth}
  \centering
  \centerline{\includegraphics[width=\textwidth]{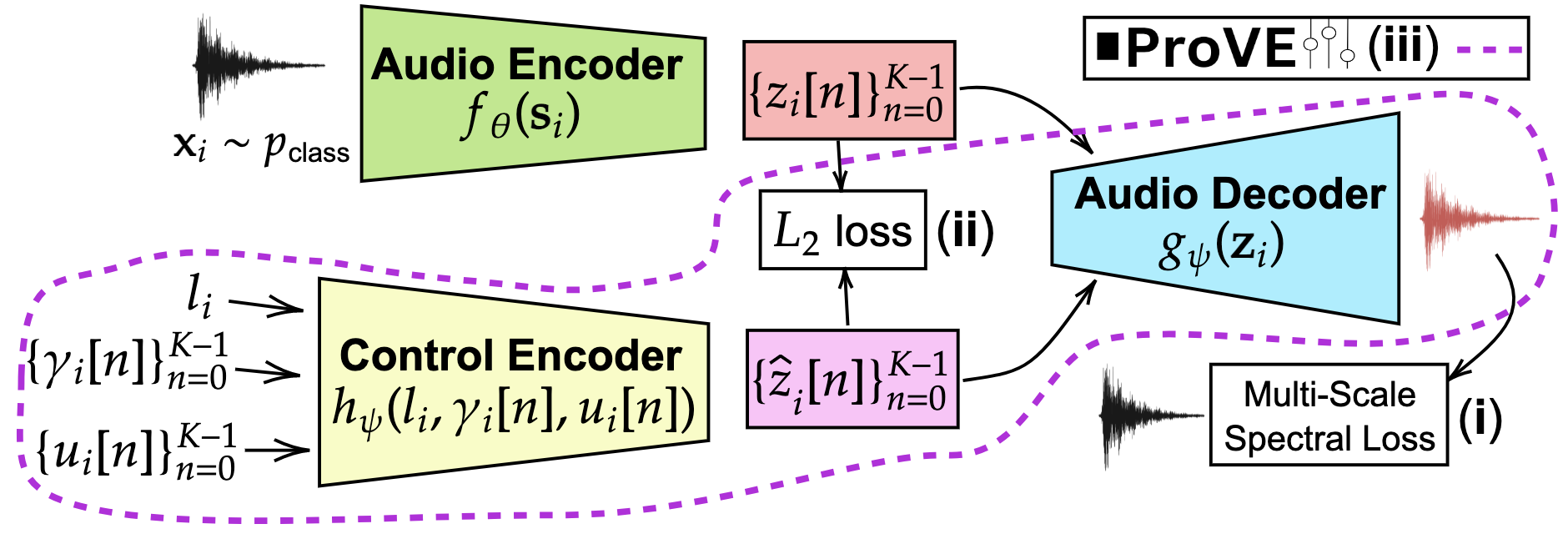}}
\end{minipage}
\caption{(\textbf{i}) and (\textbf{ii}) depict the two-stage framework of \textbf{ProVE}, where (\textbf{i}) trains a class-specific autoencoder on the audio dataset $\textbf{s}_i \sim p_{class}$, and (\textbf{ii}) trains a control parameter encoder with static condition label, temporal control proxy derived from audio $\boldsymbol{\hat{\gamma}}_i = C\{\textbf{s}_i\}$, and noise sequence. (\textbf{iii}) is the final \textbf{ProVE} model, synthesizing audio from control parameters using the trained control encoder and audio decoder.}
\label{figres}
\end{figure}

\section{ProVE Framework}
\label{secmethod}
\subsection{Overview}
\begin{figure}[htbp]
\begin{minipage}[b]{1.0\linewidth}
  \centering
  \centerline{\includegraphics[width=8.2cm]{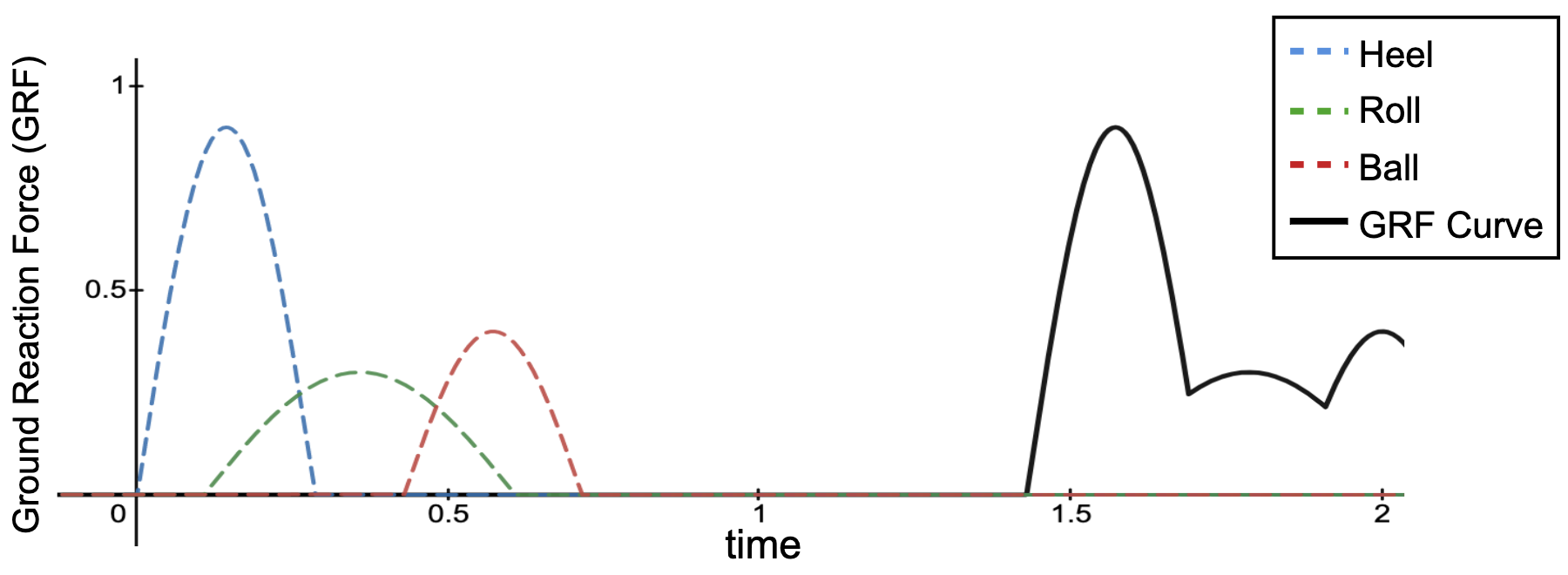}}
\end{minipage}
\caption{Sample curve from Farnell's GRF curve generator \cite{farnell2007marching}.}
\label{figres2}
\end{figure}

\textbf{PA} models generate sound from static and temporal control variables, usually accompanied by random noise. An example is the classical \textbf{PA} model for footsteps developed by Andy Farnell in \cite{farnell2007marching}. At the core of Farnell's model is a set of texture synthesis modules, one for each surface. A static surface selection control selects which texture synthesis module is active, and a ground reaction force (GRF) curve generator temporally controls the active texture synthesis module to transform uniform noise into footstep sounds. Farnell developed the GRF curve generator through the analysis of bipedal locomotion. The generator is a function of 3 polynomial segments that repeats its values over fixed intervals. The 3 polynomial segments describe ground reaction force for the phases of a human's footstep, i.e., heel, roll from heel to ball, and ball, resulting in an interpretable synthesis control signal (see Figure~\ref{figres2}).

We propose the \textbf{Pro}cedural (audio) \textbf{V}ariational auto\textbf{E}ncoder (\textbf{ProVE}), a two-stage analysis-synthesis approach to learning \textbf{PA} models which aims to improve fidelity, while retaining the properties of being controllable and expressive. This approach uses a neural audio synthesis decoder comprised of differentiable digital signal processing (\textbf{DDSP}) layers \cite{engel2020ddsp1}. More specifically, we synthesize an environmental sound class with a differentiable filtered noise module, governed by a control signal obtained through analysis, as opposed to the models that predict Fourier coefficients in the spectral domain \cite{wang2017tacotron, engel2019gansynth}, autoregressive time-domain approaches \cite{kalchbrenner2018efficient, oord2016wavenet}, as well as previously mentioned adversarial approaches, as they are typically unstable to train \cite{saxena2020generative}.

Given the highly variable nature of environmental sounds, \textbf{ProVE} aims to learn a controlled stochastic process. Here, $s[n] = \{s_n[m]\}_{m=0}^{\delta-1}$ is a small window of discrete audio of length $\delta$, sampled from:
\begin{equation} \label{eq1}
s[n] \sim q(l, \gamma[n], u[n])
\end{equation}
where $q$ is a function of sampled stochastic noise ($u[n] \sim \mathcal{U}{[-1,1]}^{n_\gamma}$) dependent on static ($l \in \mathbb{R}^{n_l}$) and temporal ($\gamma[n] \in \mathbb{R}^{n_\gamma}$) deterministic control. Let $p_{class}$ denote the prior of discrete signals of some sound class of length $L$. When we sample a discrete signal through overlapping and adding a sequence of sampled small windows $\textbf{s} = \text{OLA}(\{s[n]\}_{n=0}^{K-1})$ where $K \geq \frac{L}{\delta}$, ideally we would want $\textbf{s} \sim p_{class}$. This is opposed to models that directly output a sample array of fixed length $L$ from a static condition label, like the adversarial approach, and allows us to define our own instances of control with $l$ and $\boldsymbol{\gamma} = \{\gamma[n]\}_{n=0}^{K-1}$.

The \textbf{ProVE} framework comprises two steps, the first of which is to learn a latent audio representation space which is viewed as ``machine-interpretable'' control space, and the second is to learn a mapping from ``human-interpretable''  control (i.e. GRF Curves with surface selection) to its corresponding ``machine-interpretable'' control. Our training dataset, indexed by $i$, regardless of sound class, is formatted as 4-tuples: $(\textbf{s}_i,l_i,\boldsymbol{\hat{\gamma}}_i,\textbf{u}_i)$. $\textbf{s}_i$ is a discrete-time audio signal from a target sound class ($\textbf{s}_i \sim p_{class}$) of length $L$, $l_i$ is a static condition label associated with $\textbf{s}_i$, and $\boldsymbol{\hat{\gamma}}_i$ is an approximate temporal control variable that is proxied through signal analysis on audio $\textbf{s}_i$. That is, $\boldsymbol{\hat{\gamma}}_i = \{\hat{\gamma}_i[n]\}_{n=0}^{K-1}$ is a proxy to the actual $\boldsymbol{\gamma}_i$ which governs the temporal context of audio $\textbf{s}_i$, derived from prior knowledge of $p_{class}$. Finally, $\textbf{u}_i$ is a sequence of random noise assumed to be assigned to $\textbf{s}_i$.

\subsection{Step 1: Audio Class Representation}
Step 1 is to train an autoencoder model to learn a latent sequence representation $\textbf{z}_i = \{z_i[n]\}_{n=0}^{K-1}$ of audio $\textbf{s}_i \sim p_{class}$. Since this is a general framework, both the encoder $\textbf{z}_i = f_\theta(\textbf{s}_i)$ and decoder $\hat{\textbf{s}}_i = g_\phi(\textbf{z}_i)$ models are architecture-agnostic. However, in our experiments described in Section~\ref{secexperiments} we use the encoder and decoder models from the \textbf{DDSP} library by magenta \cite{engel2020ddsp1} for experiments due to its success in learning expressive and controllable generative models of speech \cite{fabbro2020speech} and musical instruments \cite{wu2021midi}, and its ability to synthesize high-quality audio with relatively good fidelity using smaller datasets. A trained autoencoder learns a latent sequence space that one can think of as a ``machine-interpretable'' control space to the synthesis module. The autoencoder is optimized using multi-scale spectral loss as mentioned in \cite{engel2020ddsp1}.

\subsection{Step 2: Control Mapping}
Because the latent control sequence space from step 1 is not ``human-interpretable'', a second step is needed to allow for a controllable model. In this step, a control encoder $\hat{z}_i[n] = h_\psi(l_i,\hat{\gamma}_i[n],u_i[n])$ is trained to approximate the ``machine-interpretable'' control sequence $\textbf{z}_i$, from the tuple $(l_i,\boldsymbol{\hat{\gamma}}_i,\textbf{u}_i)$ by optimizing $h_\psi$ on $L_2$-loss between $\textbf{z}_i$ and $\boldsymbol{\hat{z}}_i = \{\hat{z}_i[n]\}_{n=0}^{K-1}$ . After training, we compose the sythesis model $q(l, \gamma[n], u[n]) = (g_\phi \circ h_\psi)(l, \gamma[n], u[n])$ from the control encoder $h_\psi$ and the audio decoder $g_\phi$.

A limitation of current ProVE framework is the requirement for a simple control variable proxy $\boldsymbol{\hat{\gamma}}_i = C\{\textbf{s}_i\}$ derived from some  analysis procedure $C$ on audio $\textbf{s}_i$. In future work, we aim to mitigate this limitation by developing a procedure $C$ from existing controllable models targeting $p_{class}$, or without prior knowledge of $p_{class}$ directly from $\textbf{s}_i$.

\section{Experiments}
\label{secexperiments}
In this section we demonstrate the training of a \textbf{ProVE} model for footstep sound effects, and we compare objectively and subjectively to baselines of Farnell's classical \textbf{PA} model, as well as the GAN approach in \cite{comunita2021neural}. An interactive web application \footnote{https://dependanz.github.io/projects/prove.html} is supplemented where users can provide control variables to synthesize from a pretrained \textbf{ProVE} model of footstep sounds.

The \textbf{ProVE} models trained on the footstep sound effect datasets have a static control for the selection of surface material like in the adversarial approach. The temporal control variables $\boldsymbol{\gamma_i}$ are the GRF curves, and its approximation $\boldsymbol{\hat{\gamma}}_i$ was defined to be the smoothed envelope of the footstep sounds. That is, $\boldsymbol{\hat{\gamma}}_i = C\{\textbf{s}_i\}$ where $C$ is the 1D Laplacian smoothing of the magnitude of the discrete Hilbert transform.

\begin{figure}[htbp]
\vspace*{-0.2cm}
\begin{minipage}[b]{1.0\linewidth}
  \centering
  \centerline{\includegraphics[width=8.6cm]{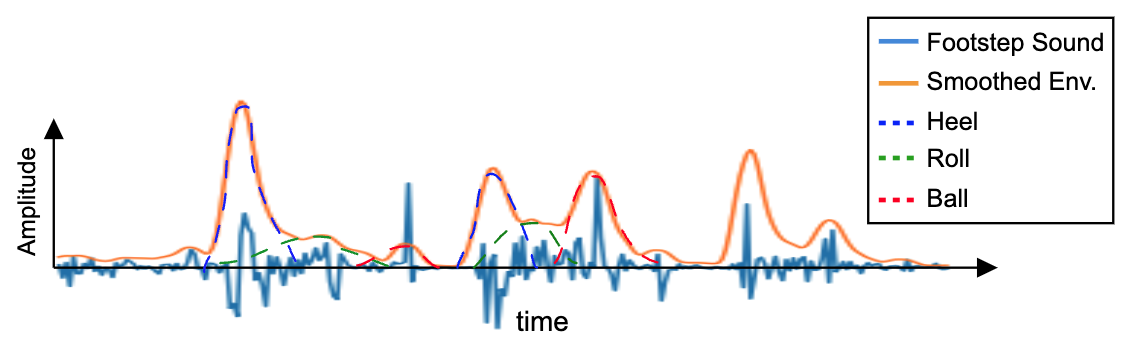}}
\end{minipage}
\vspace*{-0.75cm}
\caption{Smoothed envelope of footstep sounds.}
\label{figres3}
\end{figure}

The motivation behind the choice of proxy was that the magnitude of the exertion of force influences the magnitude of the conversion of energy which influences the loudness of an impact-based sound. The control proxy is also shown to have a 3-polynomial structure for each individual footstep that occurs in $x_i$, as can be seen in figure \ref{figres3}. 

\subsection{Datasets}
We train and evaluate \textbf{ProVE} models for footstep sounds on two publicly available datasets. The first is a set of high-heel footstep sounds from the Zapsplat\footnote{https://www.zapsplat.com/sound-effect-packs/footsteps-in-high-heels} website. This is the same dataset used in \cite{comunita2021neural}, and we use it to compare ProVE to the adversarial approach presented in \cite{comunita2021neural}. The dataset contains one-shot sounds (~1 sec) and is categorized by ground material, however, the demonstrations in the evaluation section show a ProVE model which was trained on long segments of footsteps walking on a specific material (~10 seconds) from the same dataset. The reason to train with longer audio segments instead of one-shots is that a single footstep usually doesn't occur in isolation but rather in a sequence of footsteps that has temporal structure. Thus, to improve the accuracy of PA models, we consider longer audio segments that capture this temporal structure. Bipedal acceleration sound perceptually different than bipedal deceleration, and that is due to how the control variables (GRF Curves), as well as its corresponding sound, behave over time.

The other dataset is from \textit{PremiumBeat.com}\footnote{https://www.premiumbeat.com/blog/40-free-footstep-foley-sound-effects/} and is a set of long recordings ($\sim$15-20 secs) classified by surface material and locomotion speed (Walking, Running, etc). We use this dataset to test the ability of \textbf{\textit{ProVE}} in learning higher fidelity synthesis while also retaining the same expressivity of classical \textbf{PA}. 

\subsection{Training}
For training, we resample the input audio to $f_s = 16$ kHz sample rate for a fair comparison to the baseline models, but DDSP is easily scalable to higher sampling rates. We use a $f_c = 250$ Hz control rate for $\textbf{z}_i$ and $(\boldsymbol{\gamma}_i, \textbf{u}_i)$. The main audio encoder $\textbf{z}_i = f_\theta(\textbf{s}_i)$ used for training and evaluation is a recurrent neural network from the paper that proposed \textbf{DDSP} \cite{engel2020ddsp1}, which processes normalized mel-frequency cepstral coefficients (MFCCs). The first 13 normalized MFCCs are then passed to a gated recurrent unit (GRU), whose outputs pass through a final linear layer with a $\textbf{tanh}$ activation to obtain latent control sequence $\textbf{z}_i$, where $z_i[n] \in (-1,1)^{n_z}$ with a choice of $n_z = 512$. The audio decoder $g_\theta(\textbf{z}_i)$ is a recurrent neural network which parameterizes the DDSP subtractive noise synthesizer. The control encoder $\hat{z}_i[n] = h_\psi(l_i,\hat{\gamma}_i[n],u_i[n])$ is similar to the audio encoder, in that ($\boldsymbol{\gamma}_i$, $\textbf{u}_i$) is normalized, broadcast-multiplied by a learnable embedding obtained with $l_i$, passed through a GRU and a final linear layer with the $\textbf{tanh}$ activation, giving us $\hat{z}_i[n] \in (-1,1)^{n_z}$

During inference, a \textbf{ProVE} footstep sound model is given by $\boldsymbol{\hat{s}} = \text{OLA}(\{\hat{s}[n]\}_{n=0}^{K-1})$ where $\hat{s}[n] \sim q(l, \gamma[n], u[n]) = (g_\phi \circ h_\psi)(l, \gamma[n], u[n])$. The model is controlled by providing: (i) surface material condition label $l$, (ii) a GRF curve (e.g., a sample from Farnell's GRF curve generator) as the temporal control $\boldsymbol{\gamma}$, (iii) a sample sequence of random noise $\textbf{u}$.

\subsection{Objective Evaluation} 
We use data-driven and perceptually motivated metrics Frechét Audio Distance (FAD) \cite{kilgour2018frechet} and Maximum Mean Discrepancy (MMD) \cite{gretton2012kernel} to compare the synthesis from the \textbf{ProVE} model of footstep sounds to classical \textbf{PA} models and the recent adversarial approach in \cite{comunita2021neural} (denoted \textbf{GAN}). Both of these metrics calculate a distance between audio distributions. To evaluate the models, we compute FAD and MMD between the model outputs, the training set of high-heel sounds (\textbf{Heels}) from the Zapsplat website, and the rest of the footstep sounds from the same website (\textbf{Misc}). As the quality of synthesized audio (\textbf{GAN}, \textbf{PA}, \textbf{ProVE}) increases, the distribution distance to real audio datasets (\textbf{Heels}, \textbf{Misc}) should decrease. In order to compare to the results of \textbf{GAN}, we strictly followed the same sampling and evaluation processes of \cite{comunita2021neural}, where 120k one second samples from Farnell's GRF curve were used to sample from the \textbf{PA} and \textbf{ProVE} models. With the generated footstep sounds of both models, we calculate FAD and MMD.

\begin{figure}[htbp]
\begin{minipage}[b]{\linewidth}
  \centering
  \centerline{\includegraphics[width=9cm]{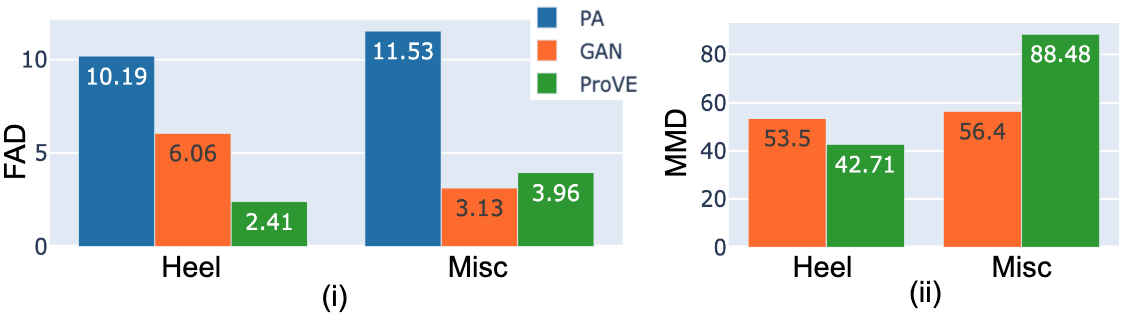}}
\end{minipage}
\vspace*{-0.75cm}
\caption{(i) Frechét Audio Distances and (ii) Maximum Mean Discrepancies of Real and Synthesized Footstep Sound Datasets}
\label{figobj}

\end{figure}

The main results in Figure \ref{figobj} show that the classical \textbf{PA} model is the furthest away from both \textbf{Heels} and \textbf{Misc}. 
The results of calculating FAD also show the \textbf{ProVE} model is closer to the training data (\textbf{Heels}) than the \textbf{GAN} model is, with an FAD of 2.41. However, the FAD between the two models and \textbf{Misc} shows \textbf{GAN} is closer with an FAD of 3.13 whereas the ProVE model has an FAD 3.96. The results on MMD are similar to FAD, with the \textbf{ProVE} model closer to \textbf{Heels} (42.71) and \textbf{GAN} further away (53.5).

These findings suggest that the \textbf{ProVE} model surpasses the \textbf{PA} model in terms of fidelity, where FAD and MMD of the \textbf{ProVE} model are significantly closer to both datasets, enforcing that the \textbf{PA} model lacks realism. This also shows that the ProVE model has a consistent distance in both datasets, where the model is closer to the training \textbf{Heels} dataset, and is proportionally further away to the out-of-distribution dataset \textbf{Misc}. However, the FAD of the GAN and ProVE models to the \textit{Misc} dataset are not dissimilar, and it can still be argued that the ProVE model also has the ability to generate samples beyond the training set.

\subsection{Subjective Evaluation}
We conducted subjective listening tests to compare the \textbf{ProVE} model, the classical \textbf{PA} model, and the training dataset, which was the set of long recordings from the PremiumBeat website (denoted PB in table \ref{tableRESEVALPROVEREAL}). We input the smoothed envelope of 8 second snippets of these recordings as the GRF curves to the \textbf{ProVE} and \textbf{PA} models. This was done in order to align the synthesized samples with real recordings for a listening test. We used a modification of pairwise tests in the Reproducible Subjective Evaluation (\textit{reseval}) framework \cite{morrison2022reproducible} such that participants were asked to choose which of two footsteps sounds more closely matched a descriptive textural prompt. We recruited 50 participants using Amazon Mechanical Turk with \textit{reseval}, and each participant completed 10 pairwise comparisons between the synthesized and real distributions. We compensated participants with a base payment and a bonus for completing the entire survey. Tables~ \ref{tableRESEVALPROVEFARNELL} and \ref{tableRESEVALPROVEREAL} display the proportion of participants that preferred \textbf{ProVE} over \textbf{PA} or the real dataset (in Table \ref{tableRESEVALPROVEREAL}), where the p-values are from two-sided binomial tests.

\input{tables/table_subjective_farnell.tex}
\input{tables/table_subjective_real.tex}

The results indicate that ProVE and classical PA models differ significantly by surface material, while comparison between ProVE and the real dataset are more similar. Participants preferred the synthesis of \textbf{ProVE} over classical \textbf{PA} for dirt, grass, and wood, but they preferred classical \textbf{PA} over ProVE for gravel. This is possibly due to the fine-grained sounds of the trajectory of small pebbles and rocks from gravel, which is not accounted for by simply using a control variable proxy of the smoothed envelope of footstep sounds.

\section{Conclusion}
\label{secconclusion}
The Procedural (audio) Variational autoEncoder (\textbf{ProVE}) was proposed as a general framework for learning high-fidelity procedural audio models of environmental sounds. \textbf{ProVE} models map from a ``human-interpretable'' to ``machine-interpretable'' control space learned from an audio autoencoder and surpass classical \textbf{PA} models in terms of fidelity while retaining controllability as a controlled stochastic process. Possible research directions from here include shortening inference time for real time audio synthesis, as well as using \textbf{ProVE} for automatic Foley generation from video/animation. The implications from \textbf{ProVE} extend beyond audio signals to procedural signal generation in general, as a step towards controllability in latent spaces.

\vfill\pagebreak

\bibliographystyle{IEEEbib}
\bibliography{paper}

\end{document}

%% file: tables/table_subjective_farnell.tex
\renewcommand{\baselinestretch}{0.8}
\let\PBS=\PreserveBackslash
\begin{table}[htbp]
\centering
\def\arraystretch{1.2}
\resizebox{7cm}{!}{%
\begin{tabular}{l|l|l|l|l|}
\cline{2-5}
p-value = $6 \times 10^{-7}$          & Dirt & Grass & Gravel & Wood \\ \hline
\multicolumn{1}{|l|}{ProVE} & \textbf{0.77} & \textbf{0.705} & 0.371  & \textbf{0.81} \\ \hline
\multicolumn{1}{|l|}{PA}    & 0.23 & 0.295 & \textbf{0.629}  & 0.19 \\ \hline
\end{tabular}%
}
\caption[Results of Subjective Evaluation (\textbf{ProVE}, \textbf{PA})]{Results of Subjective Evaluation (ProVE, PA)}
\label{tableRESEVALPROVEFARNELL}
\end{table}

\renewcommand{\baselinestretch}{1.65}

%% file: tables/table_subjective_real.tex
\renewcommand{\baselinestretch}{0.8}
\let\PBS=\PreserveBackslash
\begin{table}[htbp]
\centering
\def\arraystretch{1.2}
\begin{tabular}{l|l|l|l|l|}
\cline{2-5}
p-value = $0.1$          & Dirt & Grass & Gravel & Wood \\ \hline
\multicolumn{1}{|l|}{ProVE} & 0.5 & 0.486 & 0.431 & 0.387 \\ \hline
\multicolumn{1}{|l|}{Real (PB)}    & 0.5 & 0.514 & 0.569 & 0.613 \\ \hline
\end{tabular}
\caption[Results of Subjective Evaluation (ProVE, Real)]{Results of Subjective Evaluation (ProVE, PB)}
\label{tableRESEVALPROVEREAL}
\end{table}

\renewcommand{\baselinestretch}{1.65}